\documentstyle[multicol,prl,aps]{revtex}
\begin{document}
\def\mx{m_x}
\def\my{m_y}
\def\mz{m_z}
\def\kf{k_{\rm F}}
\def\kp{k_{\parallel}}
\def\tan  {{\rm tan}}
\def\sin  {{\rm sin}}
\def\cos  {{\rm cos}}
\def\sdag{s^\dagger}

\title
{
Angular Dependent Magnetoresistance Oscillation of a Quasi-Two-Dimensional
System in a Periodic Potential
}

\author{Daijiro {YOSHIOKA}}

\address{
Institute of Physics, University of Tokyo, Komaba, Meguro-ku,
Tokyo 153, Japan\\
}

\maketitle
\begin{abstract}
(BEDT-TTF)$_2$MHg(SCN)$_4$[M:K,Rb,Tl] shows typical two-dimensional
angular dependent magnetoresistance oscillation (ADMRO)
at high temperature (T$>$8K),
but at lower temperature it shows anomalously large magnetoresistance,
and the ADMRO pattern changes.
These low temperature behaviors are explained as effects of
a periodic potential.
The present explanation is different from that by Kartsovnik et al.
[J. Phys. I France {\bf 3} (1993) 1187]
in that reconstruction of the cylindrical Fermi surface into
an open Fermi surface is not assumed.
It is also predicted that if the periodic potential exists at quantizing
magnetic field, resistivity peak of new origin should be observed.

\end{abstract}

\begin{multicols}{2}

Quasi-two-dimensional electron system is realized in
organic conductors.
In a tilted magnetic field cylindrical Fermi surface gives rise to
the angular dependent magneto oscillation
(ADMRO).\cite{kart1,kajita,yamaji,yagi}
However, (BEDT-TTF)$_2$MHg(SCN)$_4$ [M:K,Rb,Tl] shows this kind
of oscillation only at
temperature higher than about $T_A \simeq 8$K.\cite{kago,kart3}
At lower temperature the oscillation changes into that
quite similar to that of a
quasi-one-dimensional electron system.\cite{osada}
It has also been pointed out that in the low temperature phase
magnetoresistance becomes anomalously large.\cite{osada2,sasaki1}

The change of the behavior has been attributed
 to a phase transition at $T_A$.
This series of materials has two kinds of Fermi surfaces.\cite{mori}
One is cylindrical, which causes the ADMRO,
and the other is almost flat open Fermi surface.
The latter Fermi surface causes
SDW transition at $T_A$.\cite{sasaki2,sasaki3,brooks}
Kartsovnik\cite{kart2} proposed that the new periodicity caused by the SDW
induces reconstruction of the two dimensional Fermi surface
into one-dimensional one;
the new ADMRO comes from the one-dimensional Fermi surface thus formed.

Here, we will show that the experiment can be explained without assuming
the formation of the one-dimensional Fermi surface.
We only assume that in the low temperature phase there is a periodic
potential by the SDW state.
Namely we consider a
quasi-two-dimensional system in both tilted magnetic field and the periodic
potential by the SDW modulation.

When the magnetic field is applied to a quasi-2d
system, the electron motion is restricted into a direction parallel
to the field.
Thus the system looks like one-dimensional.
In this case the effective hopping amplitude parallel to the field depends
on the direction of the magnetic field.
The periodic modulation of the hopping amplitude
as a function of the tilt angle
is the origin of the ADMRO at high temperature phase where there is no SDW.
We show that this one-dimensional system is further affected by the
periodic potential in the quasi-two-dimensional plane.
Namely the effect of the in-plane periodic potential
is converted into a periodic potential in the direction parallel to
the tilted magnetic field,
the period depending on the tilt angle.
When the period is incommensurate with the lattice periodicity in the
$z$-direction,
the electron wave function is localized.
This is the origin of the anomalously large magnetoresistance.
On the other hand when the period of the potential is integer times
that of the lattice, the wave function is extended.
This brings on the ADMRO dips in the low temperature phase.

We consider a model layered system.
The layers are parallel to the $xy$-plane and stacked in the $z$-direction.
The motion in the $xy$-plane is free,
except for the presence of the periodic potential due to the SDW.
There is a weak hopping between
adjacent layers with transfer integral $t$.
Thus the Fermi surface is a cylinder periodically modulated
in the $z$-direction,
a typical situation where two-dimensional ADMRO is observed.
A strong magnetic field is applied with angle $\theta$ from the $z$-axis
towards the $x$-axis.
In the $xy$-plane there is a uniaxial periodic potential.
Thus our Hamiltonian is
\begin{eqnarray}
H &= &{{1}\over{2m}} [p_x^2 + (p_y -eB_zx + eB_xz)^2] + {{p_z^2}\over{2m_z}}
\nonumber \\
&+ &V(z) + V_0\cos ({\bf Qr}).
\end{eqnarray}
Here $V(z)= \sum_j V_z(z-z_j)$  is the potential which confines electrons
in each layer, and the last term is the periodic potential.

In our model the effect of the periodic crystal lattice
in the $xy$-plane is represented by
effective mass in the $xy$-direction.
In general the effective mass can be anisotropic.
However, such anisotropy could easily be taken into account by scale
transformation in the $xy$-plane.
In such a transformation the direction of the magnetic field ${\bf B}$ and
${\bf Q}$ changes.
The direction of ${\bf B}$ and ${\bf Q}$ we consider here is those after the
scale transformation.
The original direction can be calculated easily.
Thus
assumption of the isotropic effective mass in the $xy$-plane causes no
loss in the general applicability of the present theory.

In this model we fix the magnetic field in the $xz$-plane, and
consider the direction of ${\bf Q}$ arbitrary:
${\bf Q} = (Q_x,Q_y,Q_z)=(Q_\bot\cos\phi,Q_\bot\sin\phi,Q_z)$.
The $z$-component of the periodic potential $Q_z$ represents relative shift
of the planar periodic potential between the layers.\footnote{
Experimentally, on the other hand, the direction of the periodic potential is
fixed relative to the crystal axis, and the direction of the magnetic field
is arbitrary.
However, isotropy in the $xy$-plane guarantees that the present choice is
general enough.}

Now we try to get eigenstates of this Hamiltonian.
When we neglect inter-layer hopping
and the periodic potential, the wave function in the $j$-th
layer is given as follows:
\begin{eqnarray}
\Phi_{n,X_j,j}({\bf r}) &= &C_n \exp{[{\rm i}k_yy
-\frac{1}{2\ell^2}(x-X_j)^2]}\nonumber \\
&\times & H_n(\frac{x-X_j}{\ell})f_0(z-z_j),
\end{eqnarray}
where $n$ is the Landau quantum number,
\begin{equation}
X_j = k_y\ell^2 + ({B_x}/B_z)z_j,
\label{center}
\end{equation}
is the center coordinate,
$z_j =cj$ is the $z$-coordinate of the $j$-th layer,
$c$ being the lattice constant,
$\ell = \sqrt{\hbar/(eB_z)}$,
$H_n$ is the Hermite polynomial,
and $C_n$  is the
normalization constant.
In the presence of the lateral component $B_x$ the center
coordinate, eq.\ (\ref{center}), depends both on $k_y$ and $z_j$.
This dependence plays an essential role in our theory.
The function $f_0(z)$ is the normalized wave function in the $z$-direction for
an isolated plane:
\begin{equation}
[\frac{1}{2\mz}p_z^2 + V_z(z)] f_0(z) = E_0 f_0(z).
\end{equation}
The transfer integral $t$ at $B=0$ is given by this function as
\begin{equation}
t = \int {\rm d}z f_0^*(z+jc)[V(z) -V_z(z)]f_0(z).
\end{equation}

In the presence of the periodic potential and the tilted field
 we cannot get exact wave functions.
Thus we consider these effects perturbationally.
As an unperturbed state we choose
\begin{equation}
\Psi_{n,k_y,k_z}({\bf r}) = \frac{1}{\sqrt{N_z}}\sum_j \exp[{\rm i}k_z z_j]
\Phi_{n,X_j,j}({\bf r}),
\end{equation}
where $N_z$ is the number of layers.
In the absence of the periodic potential
the momenta $k_y$ and $k_z$ are good quantum numbers, and
the matrix element of the
Hamiltonian is given by
\begin{eqnarray}
&&\langle \Psi_{n,k_y,k_z} | H |\Psi_{n',k_y,k_z}\rangle
= [(n+ \frac{1}{2})\hbar \omega_c + E_0] \delta_{n,n'} \nonumber \\
{}~~&&+\sum_{\tau = \pm 1} {\tilde t}_{\tau,n,n'}\exp[{\rm i}\tau k_z c],
\label{matel}
\end{eqnarray}
where ${\tilde t}_{\tau,n,n'}$ is an effective hopping amplitude,
\begin{eqnarray}
&&\tilde t_{\tau,n,n'} = t ({{n'!}\over{n!}})^{1/2} \exp[-{{1}\over{4}}
({{B_x c}\over{B_z\ell}})^2)]\nonumber \\
{}~~&\times&(-{{\tau}\over{\sqrt{2}}}~{{B_x c}\over{B_z \ell}}
)^{n'-n} L_n^{n'-n}({{1}\over{2}}~{{B_x^2c^2}\over{B_z^2\ell^2}}).
\end{eqnarray}
Here $n \ge n'$ is assumed without loss of generality,
and $L_n^m(x)$ is the Laguerre function.
In the experimental situation of $n\simeq n' \gg 1$, this effective
hopping integral $\tilde t_{j,n,n'}$ can be approximated as
\begin{eqnarray}
\tilde t_{\tau,n,n'} &\simeq &t (-\tau)^{n-n'}
({{2}\over{\pi k_{\rm F}c\tan\theta}})^{1/2}\nonumber \\
&\times& \cos[\kf c \tan\theta -
{{n'-n}\over{2}}\pi - {\pi\over4}],
\end{eqnarray}
where $\kf$ is the two-dimensional Fermi momentum.
Therefore $\tilde t_{j,n,n}$ vanishes periodically with respect to
$\tan{\theta}$.
This gives peaks in the $z$-axis resistivity.
Essentially similar calculation has been done,\cite{osada1,kuri} and the
results coincide with semiclassical treatment of the
ADMRO.\cite{yamaji,yagi}
In this quantum mechanical treatment
the oscillation is a consequence of interference of wave functions
when an electron hops between layers:
Conservation of $k_y$ means shift of center coordinate as a course of
hopping.
Therefore integration with respect to $x$ of product of oscillating
Hermite polynomials causes interference.
Equation\ (\ref{matel}) shows that the system looks
like a one-dimensional system
with electron motion parallel to the direction of the magnetic field.

On the other hand, once the periodic potential
$V_0\cos ({\bf Qr})$ is turned on,
mixing of the wave numbers $k_y$ and $k_z$ occurs:
The matrix element of the periodic potential is given as
\begin{eqnarray}
&&\langle \Psi_{n,k_y,k_z} | V_0\cos (Qx) | \Psi_{n',k_y',k_z'}\rangle
= {{1}\over{2}} \sum_{\tau=\pm}
\nonumber \\
& &V^{\rm eff}_{n,n',k_y,k_z}(\tau Q_x,\tau Q_y)
 \delta_{k_y,k_y'+\tau Q_y}
 \delta_{k_z,k_z'+\tau (Q_z+Q_xB_x/B_z)},
\end{eqnarray}
where
\begin{eqnarray}
&&V^{\rm eff}_{n,n',k_y,k_z}(Q_x,Q_y)
=
\sqrt{{{n'!}\over{n!}}}[{{{\rm i}\ell}\over{\sqrt{2}}} (Q_x + i Q_y)]^{n-n'}
\nonumber \\
&&\times \exp[-{1\over 4}Q_\bot^2\ell^2 +
{\rm i} Q_x(k_y-{{Q_y}\over{2}})\ell^2]
L_{n'}^{n-n'}[{1\over 2} Q_\bot^2\ell^2],
\nonumber \\
&&\propto
\cos[\sqrt{2n}Q_\bot\ell - {{n'-n}\over{2}}\pi - {\pi \over 4}],
\end{eqnarray}
again for the case of $n \ge n'$,
and the last line is for the semiclassical case of $n\simeq n' \gg 1$.
Selection rule for $k_z$ means that there is an effective periodic potential
in the $z$-direction,
whose wave number is given by $Q_z+(B_x/B_z)Q_x = Q_z +\tan\theta\cos\phi
Q_\bot$ and strength proportional
to $L_n(Q_\bot^2\ell^2/2)$.
It should be noted that the period depends on the tilt angle
and the strength depends on  that of the magnetic field.
This mixing of $k_z$ can be understood as follows.
An electron in the $j$-th layer
at center coordinate $X$ feels the potential around $X$ averaged over
$\ell$.
When it hops between layers, the momentum $k_y$ is conserved,
thus the center coordinate shifts by $(B_x/B_z)c$.
Namely as an electron hops between layers, the local potential that electron
feels changes.
This gives rise to the periodic potential in the hopping direction, which
is parallel to the magnetic field.

Double periodicity in the $z$-direction tends to increase the $z$-axis
resistivity.
This explains the anomalously large magnetoresistance.
Two-dimensional ADMRO is buried in this increase of the resistance.
However, such effect vanishes when the condition
\begin{equation}
Q_z+\tan\theta\cos\phi Q_\bot = {{2\pi n}\over{c}},
\end{equation}
is satisfied with $n$ being an integer.
Namely dips in the resistivity are expected when $\theta$ and $\phi$
satisfies this condition.
It reproduces the experimentally observed locations of
dips,\cite{kart2,iye,kovalev,sasaki}
\begin{equation}
\tan\theta\sin(\phi - \phi_0) \simeq 0.5 + 1.25 n,
\end{equation}
if we assume $Q_\bot \simeq 0.25\AA^{-1}$,
$Q_z/Q_\bot \simeq -0.5$, and the direction of ${\bf Q}$
to be $\phi_0 \simeq 30^\circ$ from the c$^*$ axis in the a-c$^*$
plane.\footnote{In this material a-c$^*$-plane is the 2-d $xy$-plane,
and b$^*$-axis is our $z$-axis.}
This size and direction of ${\bf Q}$ is essentially the same as the proposal
by Kartsovnik et al.\cite{kart2}
Thus the present theory explains the experiments without assuming formation
of an open Fermi surface out of the cylindrical one in the low temperature
phase.\footnote{The strength of the periodic potential oscillates
with ${\bf B}$: $V^{{\rm eff}} \propto \cos[\sqrt{2n}Q_\bot\ell]$.
For $Q_\bot \simeq 0.25\AA$ this oscillation is too rapid to be observed.}

Our theory is better than the previous one\cite{kart2} in several points:
(1) Since in our theory the two dimensional Landau subband remains as it is,
it is consistent with the observation of the  Schubnikov-de Haas oscillation
in this phase.\cite{osada2,sasaki3,uji}
(2) Reconstruction of the Fermi surface depends on delicate balance between
the size of the Fermi surface and direction of the modulation potential,
so for K-salt such reconstruction may not be possible.\cite{kovalev}
(3) Explanation by the open Fermi surface requires unphysical long-ranged
transfer integral between chains
($t_{m,1}$ in eq.(3) of ref.\onlinecite{osada}).\cite{iye}

Finally we predict possible appearance of resistance peaks in higher
magnetic field.
In the magnetic field range ADMRO has been observed, there are a lot of
Landau subbands crossing the Fermi level.
However, at higher magnetic field only a few subbands exists.
In this case if $Q_z + \tan\theta\cos\phi Q_\bot$ coincides with $2\kf$ of
one of the subbands, energy gap is formed at $E_{\rm F}$, thus resistivity
becomes larger.
Since $2\kf$ depends on the strength of the magnetic field ${\bf B}$,
the peak position should depends both on the strength and direction of the
field.
Unfortunately, the SDW of (BEDT-TTF)$_2$MHg(SCN)$_4$ system
is destroyed at higher magnetic field,
so it is hard to observe this peak in this material.
However, we hope observation of the peak in other materials in near
future.

This work is supported by Grant-in-Aid for Scientific Research on Priority
Areas ``Novel Electronic States in Molecular Conductors" (Grant No.
06243103) from Ministry of Education, Science and Culture.

\end{multicols}

\end{document}